\documentclass[twocolumn,showpacs,superscriptaddress,amsmath,amssymb,nofootinbib]{revtex4}
\usepackage{graphicx}
\usepackage{dcolumn}
\usepackage{bm}

\def\simge{\mathrel{
     \rlap{\raise 0.511ex \hbox{$>$}}{\lower 0.511ex \hbox{$\sim$}}}}
\def\simle{\mathrel{
     \rlap{\raise 0.511ex \hbox{$<$}}{\lower 0.511ex \hbox{$\sim$}}}}
\def\be{\begin{equation}}
\def\ee{\end{equation}}
\def\bea{\begin{eqnarray}}
\def\eea{\end{eqnarray}}
\def\dblone{\hbox{$1\hskip -1.2pt\vrule depth 0pt height 1.6ex width 0.7pt
                  \vrule depth 0pt height 0.3pt width 0.12em$}}

\newcommand{\calf}{{\cal F}}
\newcommand{\calfmf}{{\cal F}_{\rm mf}}
\newcommand{\calz}{{\cal Z}}

\newcommand{\boldl}{{\bf L}}
\newcommand{\tr}{{\rm tr}\,}
\renewcommand{\d}{{\rm d}}
\newcommand{\cals}{{\cal S}}

\newcommand{\real}{{\rm Re\,}}

\begin{document}
\title{Eigenvalue repulsion in an effective theory of
SU(2) Wilson lines in three dimensions}

\author{Adrian Dumitru$^1$ and Dominik Smith}
\affiliation{
Institut f\"ur Theoretische Physik,
Johann Wolfgang Goethe-Universit\"at,
Max-von-Laue-Str.\ 1, 60438  Frankfurt am Main, Germany
}

\date{\today}
\begin{abstract}
We perform simulations of an effective theory of SU(2) Wilson lines in
three dimensions. We include a non-perturbative ``fuzzy-bag''
contribution which is added to the one-loop perturbative potential for
the Wilson line. We confirm that, at moderately weak coupling, this
leads to eigenvalue repulsion in a finite region above the deconfining
phase transition which shrinks in the extreme weak-coupling limit. A
non-trivial Z(N) symmetric vacuum arises in the confined phase.
\end{abstract}
\maketitle

\section{Introduction} \label{sec:Intro}
In QCD at very high temperature, the pressure is due to weakly
interacting quasi-particle gluons.  (Here, we ignore contributions
from quarks and focus on the pure gauge theory.) It can be calculated
from an effective theory in three dimensions\footnote{That is, all
fields in~(\ref{Leff_pert}) are functions of $\bm{x}$ only, and the
action is given by an integral of the Lagrangian over space, divided
by temperature.},
\be \label{Leff_pert}
{\cal L}^{\rm eff} = \frac{1}{2}\tr G^2_{ij} +\tr |D_i A_0|^2
  + m_D^2\tr A_0^2 + \cdots;
\ee
see~\cite{Pisarski:2006hz,Vuorinen:2006nz} and references
therein. $G_{ij}$ is the magnetic field strength associated with the
spatial components of the vector potential $\bm{A}$, $m_D$ is the
Debye mass, and the dots represent self-interactions of $A_0$. The
effective theory~(\ref{Leff_pert}) is valid as long as fluctuations in
$A_0$ are small; that is, the expectation value of the Polyakov loop
in the original four-dimensional theory, which is given by the trace
of the thermal Wilson line,
\bea
\ell(\bm{x}) &=& \frac{1}{N} \tr\boldl(\bm{x})~, \label{RenPL}\\
\boldl(\bm{x}) &=& {\cal Z}_R^{-1} \,
    {\cal P} \exp\left( ig\int\limits_0^{1/T} \d\tau\;
      A_0(\bm{x},\tau) \right)~,
\eea
should be close to one of the $N$ roots of unity, where $N$ is the
number of colors. We have indicated explicitly that in the
four-dimensional theory $A_0(\bm{x},\tau)$ depends on Euclidean time
$\tau$ and that Polyakov loops have to be renormalized to obtain a
non-zero continuum limit~\cite{Kaczmarek:2002mc,dhlop}. The
renormalization constant ${\cal Z}_R^{-1}$ depends on the representation
of $\boldl$, taken here to be the fundamental representation; thus,
$\boldl$ represents the propagator of an infinitely heavy test
quark. Also, we always consider normalized traces and divide by the
dimension of the representation.

The Polyakov loop represents an order parameter for the spontaneous
breaking of the global Z(N) center-symmetry corresponding to gauge
transformations which are periodic in $\tau$ only up to an element of
Z(N). In the high-temperature deconfined phase, the Polyakov loop
aquires a non-vanishing expectation value but vanishes in the confined
phase~\cite{McLSv}. As a consequence of the Z(N) symmetry in the
original four-dimensional theory, when $A_0/T$ is large (of order
$1/g$), the effective electric field in three dimensions is not simply
$E_i(\bm{x})=D_i(\bm{x}) A_0(\bm{x})$~\cite{Pisarski:2006hz,DiakoOsw}.

For two colors, the phase transition is of second
order~\cite{Engels:1994xj,Velytsky:2007gj} and so $\langle\ell\rangle
(T_d) = 0$ vanishes continuously (the theory is in the Z(2)
universality class~\cite{Svetitsky:1982gs}). Hence, in the immediate
vicinity of $T_d$ at least, the Polyakov loop is clearly far from
unity. For $N=3$, the transition is first-order~\cite{Boyd:1996bx} and
$\langle\ell\rangle$ is discontinuous at $T_d$. If the SU(3) gluon
plasma was perturbative all the way down to $T_d^+$ (in the electric
sector, $\langle\ell\rangle(T_d^+) \approx 1$), then~(\ref{Leff_pert})
{\em might} have applied even for $T$ just above $T_d$. However,
lattice measurements~\cite{Kaczmarek:2002mc,dhlop} indicate that
$\langle\ell\rangle(T_d^+) \simle0.5$, which is rather far from
unity. Moreover, the ratio of screening masses defined from two-point
correlation functions of the real and the imaginary part of the
Polyakov loop, respectively, increases from $\approx$3:2 (which is the
LO result from perturbation theory) at high temperature to $\approx$3
near $T_d$~\cite{Dumitru:2002cf}. Also, resummations of perturbation
theory work very well at high $T$ but appear to fail to reproduce the
pressure or the entropy density~\cite{Andersen:2004fp} below $\approx
3 T_d$. Finally, the interaction measure $(e-3p)/T^4$ for both
two~\cite{Engels:1994xj} and three~\cite{Boyd:1996bx} colors is rather
large up to $T\approx3 T_d$.

These observations may suggest that at temperatures not very far above
$T_d$, that even at weak coupling the theory is non-perturbative in
the sense that $A_0/T$ is large. If so, it is useful then to construct
an effective theory in terms of the Wilson line $\boldl$ rather than
$A_0$~\cite{Pisarski:2006hz,Vuorinen:2006nz,deForcrand:2008aw,
  Pisarski:2000eq}. This Lagrangian can also incorporate the global
Z(N) symmetry for the Polyakov loop.

As shown in ref.~\cite{Pisarski:2006hz}, the electric field in the
three-dimensional theory for {\em arbitrary} $A_0$ is given by
\be
E_i(\bm{x}) = \frac{T}{ig} \, \boldl^\dagger (\bm{x}) 
 D_i(\bm{x}) \boldl(\bm{x}) ~.
\ee
The classical Lagrangian in three dimensions then becomes
\be  \label{Leff_cl}
{\cal L}^{\rm eff}_{\rm cl} = \frac{1}{2}\tr G^2_{ij} + \frac{T^2}{g^2}
   \tr |\boldl^\dagger D_i \boldl|^2 ~.
\ee
Contrary to sigma models with left-right symmetry, for loops there is
also a potential. It can be written as an infinite sum over all Z(N)
neutral loops~\cite{dlp}. For the present purposes, however, we rather
write it in terms of powers of the fundamental Wilson line. To one
loop and for constant $\boldl$~\cite{GPY},
\be  \label{Veff_1loop}
{\cal L}^{\rm eff}_{\rm 1-loop} = - \frac{2}{\pi^2} T^4
  \sum\limits_{n\ge1} \frac{1}{n^4} |\tr \boldl^n|^2 ~.
\ee
This potential is evidently minimized by the perturbative vacuum
$\langle\boldl\rangle=\dblone$ (times a phase), for any $T$. To
generate a phase transition in infinite volume,
ref.~\cite{Pisarski:2006hz} suggested to add non-perturbative
contributions such as
\be  \label{Veff_fuzzyB}
{\cal L}^{\rm eff}_{\rm non-pert.} = B_f T^2 |\tr\boldl|^2~,
\ee
with $B_f$ a ``fuzzy'' bag constant (see, also,
refs.~\cite{Meisinger,Megias}). At sufficiently low temperature,
(\ref{Veff_fuzzyB}) dominates over the perturbative
potential~(\ref{Veff_1loop}) and induces a transition to a confined
phase with $\langle\tr\boldl\rangle=0$. It was further suggested
in~\cite{Pisarski:2006hz} that terms such as~(\ref{Veff_fuzzyB}) lead
to ``repulsion'' of the eigenvalues of the Wilson line in some
temperature range above $T_d$. If so, then the distribution of
eigenvalues should deviate from a sharp peak near 1 for non-asymptotic
temperatures. Our numerical results confirm this idea in the regime
where the nearest-neighbor coupling $\beta\sim 1/g^2$ is not so large
as to suppress fluctuations of the Wilson lines in space.

In this paper, we perform Monte-Carlo simulations of an effective
theory motivated by~(\ref{Leff_cl}-\ref{Veff_fuzzyB}) on a
three-dimensional lattice. The theory is defined with a spatial cutoff
on the order of the inverse temperature as~(\ref{Leff_cl}) is
non-renormalizable in three dimensions and is valid only over distance
scales larger than $1/T$~\footnote{A related renormalizable theory has
been formulated in refs.~\cite{Vuorinen:2006nz,deForcrand:2008aw}.
Ref.~\cite{Kurkela:2007dh} derived the relations between lattice and
continuum theories to leading order in lattice perturbation
theory.}. We shall focus in particular on measuring the eigenvalue
distribution both above and at the (de-)confining phase transition,
thereby testing the presence of eigenvalue repulsion in the phase
transition region. We presently employ several approximations which
simplify the simulations drastically. Most importantly, the present
simulations neglect the magnetic sector,
\be
A_i=0~.
\ee
Hence, the gauge theory is essentially reduced to a sigma model. A
precise matching of the couplings in the effective theory to
correlation functions measured in the continuum limit of the original
four-dimensional theory is beyond the scope of this
paper. Furthermore, we neglect all but the $n=1$ term
in~(\ref{Veff_1loop}), which can then be combined with the
non-perturbative potential~(\ref{Veff_fuzzyB}).

It should be noted, in particular, that the matrix model studied below
is in a different universality class than four-dimensional SU(2)
Yang-Mills theory (for a recent discussion of the latter, see
ref.~\cite{Velytsky:2007gj}). Therefore, near the transition
long-distance properties will not match. Nevertheless, we
introduce~(\ref{eq:action}) here as a simple realization of a matrix
model which allows us to study the distribution of eigenvalues of
$\boldl$ in the plane of nearest-neighbor matrix coupling $\beta$ and
``fuzzy bag'' constant (or temperature) $m^2$.

\section{The Lattice Action}  \label{sec:LattAct}

Our general three-dimensional lattice action includes kinetic
(nearest-neighbor interaction) and mass terms,
\be  \label{eq:action}
\cals = - \frac{1}{2}\beta\sum\limits_{\langle ij\rangle}
              \tr\left( \boldl_i\boldl_j^\dagger + {\rm h.c.}\right) 
        - m^2 \sum\limits_i |\tr\boldl_i|^2~, 
\ee
where $\boldl$ denotes SU(2) Wilson lines in the fundamental
representation, $i$ labels sites, and $\langle ij\rangle$ labels
links. We employ periodic boundary conditions. The kinetic term is
invariant under global ${\rm SU}_L(2) \times {\rm SU}_R(2)$
transformations while the mass term breaks it to ${\rm SU}(2)$. The
weak-coupling limit of the original four-dimensional theory
corresponds to large $\beta$. The partition function involves an
integral over the invariant SU(2) measure $[\d\boldl]$ at each site,
\be  \label{eq:Z}
\calz = \int \prod\limits_n~ [\d\boldl_n]~~e^{-\cals}~.
\ee

\section{Mean-field approximation}  \label{sec:MFA}

The mean-field approximation for the matrix model has been discussed
in detail in refs.~\cite{KSS,dhlop,dlp,Damgaard:1987wh}. We briefly
review the main steps and results as required for our present
purposes.

Replace all $2d$ nearest neighbors of any given site
in~(\ref{eq:action}) by a fixed matrix $\overline\boldl$, where $d=3$ is
the number of spatial dimensions. This defines a single-site free
energy,
\be
e^{-N_s^d \calf_{ss}(\overline\boldl)} = \calz_{ss}^{N_s^d}~,
\ee
where $N_s$ is the number of sites per spatial dimension and
\be
\calz_{ss} = \int [\d\boldl]~\exp\left[
           d\beta\; \tr\left( \boldl\overline\boldl^\dagger + {\rm h.c.}\right) 
        + m^2\; |\tr\boldl|^2~\right]~.
\ee
Consistency requires that
\be
\langle (\boldl)^*_{lk}\rangle = \frac{1}{d\beta} 
                  \frac{\partial}{\partial (\overline\boldl)_{lk}} 
		  \log \calz_{ss}(\overline\boldl)
\ee
be equal to
\be
(\overline\boldl)^*_{lk} = \frac{\partial}{\partial (\overline\boldl)_{lk}} 
		  (\overline\boldl)^*_{lk}(\overline\boldl)_{lk}~.
\ee
It follows that $\overline\boldl$ minimizes a mean-field free energy
defined as
\bea
0 &=& \frac{\partial}{\partial \overline\boldl}\; \calfmf(\overline\boldl)~, \\
\calfmf(\overline\boldl) &=& \calf_{ss}(\overline\boldl) +
  d\beta\;\tr\overline\boldl^\dagger\overline\boldl~.
\eea
To proceed, we assume that $\overline\boldl$ is proportional to the
unit matrix, $\overline\boldl=\overline\ell\,\dblone$ (for two colors,
$\overline\ell$ can be chosen to be real), so that
\be
e^{-\calf_{ss}(\overline\ell)} = 
\int [\d\boldl]~\exp\left[2d\beta\overline\ell\;\tr(\boldl +\boldl^\dagger)
                       + m^2\,|\tr\boldl|^2\right]~.
\ee
The action is a function only of the trace of the integration
variable, so that we can write 
\be  \label{eq:Lphi}
\boldl=\exp\,{\rm diag}\,(i\phi,-i\phi+2\pi i n)~,
\ee
with $n$ an arbitrary integer, and
employ Weyl's parameterization
\be \label{eq:Weyl}
[\d\boldl] \sim \d\phi \; |\Delta(\phi)|^2 = \d\phi\, \sin^2\phi~,
\ee
where $\Delta(\phi)$ denotes the Vandermonde determinant. Up to an
overall constant then,
\bea
e^{-\calf_{ss}(\overline\ell)} &=& 
\int\limits^1_{-1} \d\cos\phi\nonumber\\
& & \hspace{-1cm} \exp\left[4d\beta\overline\ell\cos\phi
              + 4m^2\cos^2\phi +
	      \frac{1}{2}\log(1-\cos^2\phi)\right]~.
   \nonumber\\
& &   \label{eq:Fss_MF}
\eea
This integral could now be evaluated analytically in a saddle-point
approximation. However, we have found that for $d=3$ the analytical
result is too inaccurate to be useful in practice, in particular in
the interesting region of $\beta$ and $m^2$. Therefore, we have rather
tabulated~(\ref{eq:Fss_MF}) as a function of $\overline\ell$. The
expectation value $\ell_0$ of $\tr\boldl/2$ is then given by the
location of the minimum of
\be
\calfmf(\overline\ell) = \calf_{ss}(\overline\ell) + 
                        2d\beta{\overline\ell}^{\,2}~.
\ee

\section{Results}  \label{sec:Results}

\subsection{The model with global ${\rm SU}_L(2)\times {\rm SU}_R(2)$ 
symmetry} 

We begin with the pure nearest-neighbor interaction model
with no loop potential, corresponding to
eqs.~(\ref{eq:action},\ref{eq:Z}) with $m^2=0$~\cite{KSS}:
\be  \label{eq:action_m2=0}
\cals = - \frac{1}{2}\beta\sum\limits_{\langle ij\rangle}
              \tr\left( \boldl_i\boldl_j^\dagger + {\rm h.c.}\right)~. 
\ee
Note that in~(\ref{eq:action_m2=0}) the basic degrees of freedom are
the Wilson line matrices, or their eigenvalues; the model therefore
differs from others which deal exclusively with the trace of $\boldl$,
such as $\cals \sim -\beta \sum \,(\tr\boldl_i \, \tr\boldl_j^\dagger
+ {\rm c.c.})$~\cite{Svetitsky:1985ye}. Alternatively, one may consider
nearest-neighbor interactions between Polyakov loops in arbitrary
representations~\cite{dhlop,dlp,heinzl}.

We expect that for small $\beta$ there is a phase where the adjoint
fields
\be  \label{eq:pion}
\tilde\ell^a(\bm{x}) = \frac{1}{2i}\tr\boldl(\bm{x})\bm{\tau^a}
\ee
as well as the singlet field (which is actually the Polyakov loop)
\be  \label{eq:sigma}
\ell(\bm{x}) = \frac{1}{2}\tr\boldl(\bm{x})
\ee
are massive\footnote{We assume hermitian generators normalized
  according to $\tr\bm{\tau^a\tau^b}=2\delta^{ab}$.}. Furthermore, the
expectation value of the ``length'' of $\overline\boldl$,
\be \label{eq:KSS_OP} 
u = \sqrt{\tr\overline\boldl^\dagger \overline\boldl/2}~~,~~
u_0 = \left< u\right> ~, 
\ee
should vanish also. The bar stands for the average over the volume for
any given configuration:
\be \label{eq:KSSorder}
\overline\boldl = \frac{1}{N_s^3}\sum\limits_i \boldl_i~,
\ee
while $\langle\cdot\rangle$ is the average over configurations. Note
that $\tr\overline\boldl^\dagger \overline\boldl /2 = 2
(\tr\overline\boldl/2)^2 -\tr \overline \boldl^2/2 \equiv \overline\ell^2 -
\overline\ell_2$, where $\ell_2$ is the Polyakov loop with Z(N) charge
two~\cite{Pisarski:2002ji} (which is neutral when $N=2$).

For sufficiently large $\beta$, on the other hand, the Wilson lines at
different sites have to align in order to minimize the
action~(\ref{eq:action_m2=0}). Hence, for a given configuration
(resp.\ Metropolis time) $\overline\boldl$ should be
non-zero. However, its direction in group space will rotate
from configuration to configuration, implying
$\langle\overline\boldl\rangle=\bm{0}$. To monitor the transition to
an ordered phase at large $\beta$ we therefore use $\langle u\rangle$
rather than $\langle \tr \overline\boldl\rangle$ as order
parameter~\cite{KSS}. Alternatively, one could add a weak background
field, $-h\,\tr\boldl$, which is then taken to zero after the
extrapolation to infinite volume has been performed.

The regimes where $\langle u\rangle =0$ and $\langle u\rangle \ne0$,
respectively, are separated by a second-order phase transition at some
critical $\beta_c$~\cite{KSS} which we determine numerically. This
transition is associated with spontaneous breaking of the ${\rm
  SU}_L(2)\times {\rm SU}_R(2)$ symmetry to ${\rm SU}_V(2)$, where
three Goldstone modes appear.

\begin{figure}
\includegraphics*[width=\linewidth]{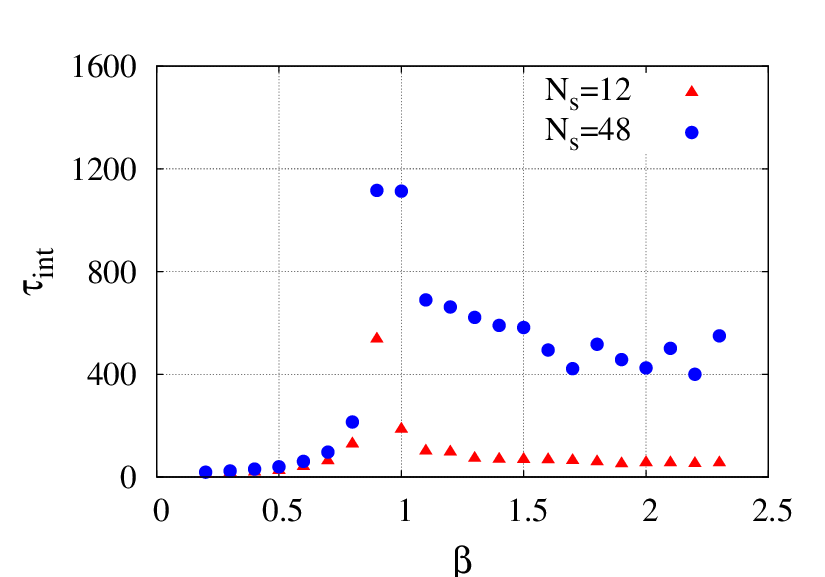}
\caption{The integrated autocorrelation time
as a function of the coupling $\beta$ for various lattices.
  \label{fig:KSS_tauint_beta}}
\end{figure}
The ensemble average denoted by $\langle\cdot\rangle$ should be
performed over statistically independent configurations. It is
therefore necessary to determine the autocorrelation time of the
Monte-Carlo algorithm as a function of $\beta$. This is done via the
``rebinning method''~\cite{berg} as follows. First, we group the
sequence ${\cal O}_i$ of measurements of a given operator\footnote{We
  take ${\cal O}=u$ defined in eq.~(\ref{eq:KSS_OP}).} into $N_{bs}$
bins of size $N_b$,
\be
{\cal O}_j(N_b) = \frac{1}{N_b} \sum\limits_{i=jN_b}^{(j+1)N_b-1} {\cal O}_i~,
\ee
where $j=0\cdots N_{bs}-1$ labels the bins. Hence, ${\cal
  O}_j(N_b)$ is simply the mean over the measurements belonging to
the bin $j$. We then determine the variance of the new sequence ${\cal
  O}_j(N_b)$:
\be
\sigma^2_{N_b} = \frac{1}{N_{bs}} \sum\limits_{j=0}^{N_{bs}-1} 
\left( {\cal O}_j(N_b) - \langle{\cal O}\rangle\right)^2~.
\ee
$\langle{\cal O}\rangle$ denotes the average of ${\cal O}$ over all
configurations.  The integrated autocorrelation time corresponding to
the bin-size $N_b$ is given by
\be
\tau_{\rm int}(N_b) = \frac{\sigma^2_{N_b}}{\sigma^2}~,
\ee
where $\sigma^2$ denotes the variance of the original sequence of
measurements. We then plot $\tau_{\rm int}(N_b)$ versus $N_b$, which
eventually approaches a flat plateau (up to rapid oscillations). This
defines $\tau_{\rm int}$, which is shown in
Fig.~\ref{fig:KSS_tauint_beta} as a function of $\beta$. In the
vicinity of the critical point, the Metropolis update algorithm
displays the well-known critical slowing down phenomenon; $\tau_{\rm
int}$ diverges in the infinite-volume limit. The measurements obtained
on $N_s=24$, 36, 48 lattices can be fitted with the form $\tau_{\rm
int} \sim N_s^{1/\nu_\tau}$, with the scaling exponent
\be
\nu_\tau = 0.72(4)~.
\ee
Away from $\beta_c$ the autocorrelation time decreases. Notice,
however, that it increases with the volume even above the critical
point, while it exhibits the standard behavior for $\beta<\beta_c$.

In practice, our simulations are performed as follows. The initial
configuration of SU(2) matrices is chosen randomly. We then perform a
number of thermalization steps which is larger than the
autocorrelation time $\tau_{\rm int}$ (determined beforehand in a
pre-run) of the order parameter. Subsequently, measurements are
performed in time intervals slightly larger than $\tau_{\rm int}$. We
employ a standard Metropolis update~\cite{berg} where all $N_s^3$
sites are scanned in sequence. Typically, we summed on the order of
thousand configurations for each set of couplings.

\begin{figure}
\includegraphics*[width=\linewidth]{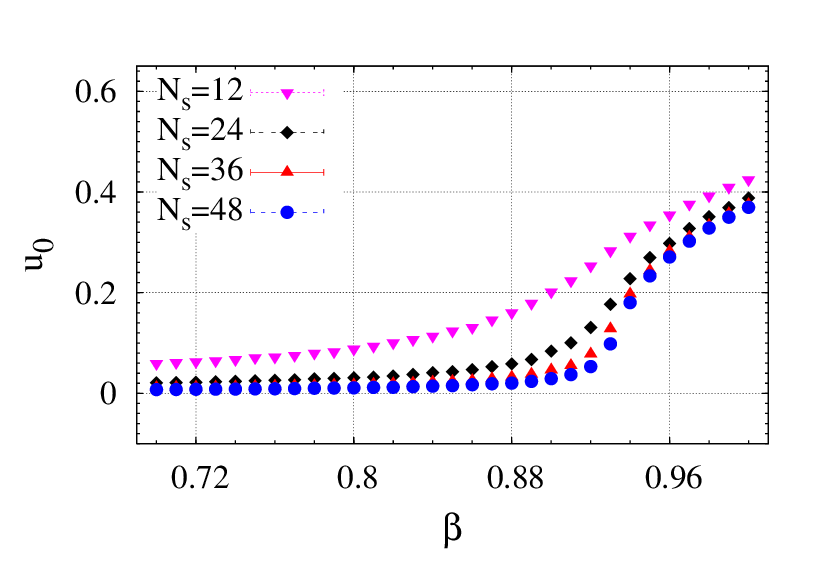}
\caption{The length of the O(4)-like order parameter from
  eq.~(\ref{eq:KSS_OP}) as a function of the coupling $\beta$ for
  various lattices.
\label{fig:KSS_ell0_beta}}
\end{figure}
Fig.~\ref{fig:KSS_ell0_beta} shows the expectation value of the order
parameter~(\ref{eq:KSS_OP}) as a function of $\beta$ on lattices of
various sizes. Statistical error bars are smaller than the size of the
symbols. There is, clearly, a order-disorder transition at
$\beta_c\simeq0.9$. As expected, finite-size effects are visible
around the transition point ($\beta\simeq \beta_c$). We have verified
that $u_0$ approaches 1 for $\beta\gg1$.

\begin{figure}
\includegraphics*[width=\linewidth]{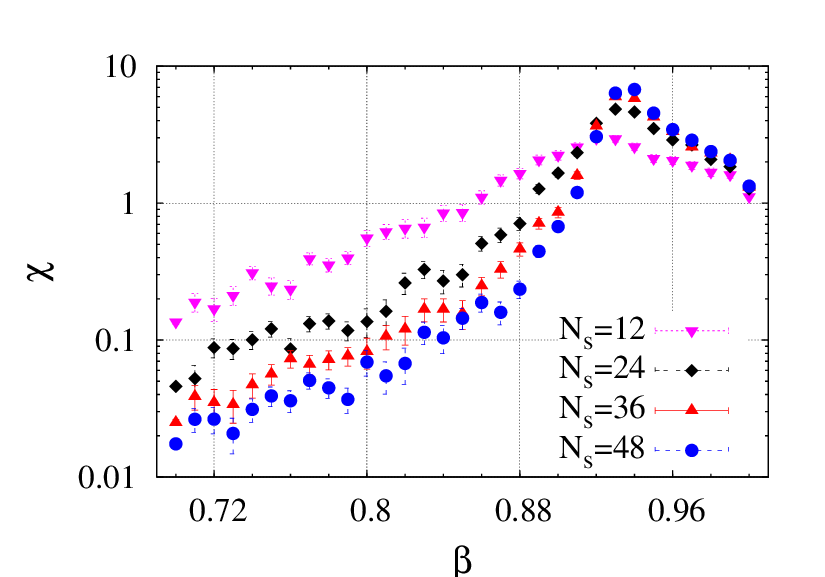}
\caption{The derivative of the order parameter from
  eq.~(\ref{eq:KSS_OP}) with respect to $\beta$ as a function of the
  coupling $\beta$ on various lattices.
\label{fig:KSS_chi_beta}}
\end{figure}
To estimate the infinite-volume limit of $\beta_c$ we proceed as
follows. We first determine the temperature susceptibility
$\chi(\beta)=\partial u_0 / \partial \beta$, as shown in
Fig.~\ref{fig:KSS_chi_beta}.
The location of the maximum defines $\beta_c$ for any given lattice
size. Extrapolating linearly to $1/N_s= 0$, we obtain
\be
\beta_c = 0.942(5)~.
\ee
We have verified that the derivative of the average kinetic energy per
link $E\sim \real\langle\tr\boldl_i^\dagger \boldl_{i+1}\rangle$ with
respect to $\beta$ also peaks at the same value of the coupling, which
is somewhat larger than the estimate from ref.~\cite{KSS}, who
employed smaller lattices and lower statistics.

\begin{figure}
\includegraphics*[width=\linewidth]{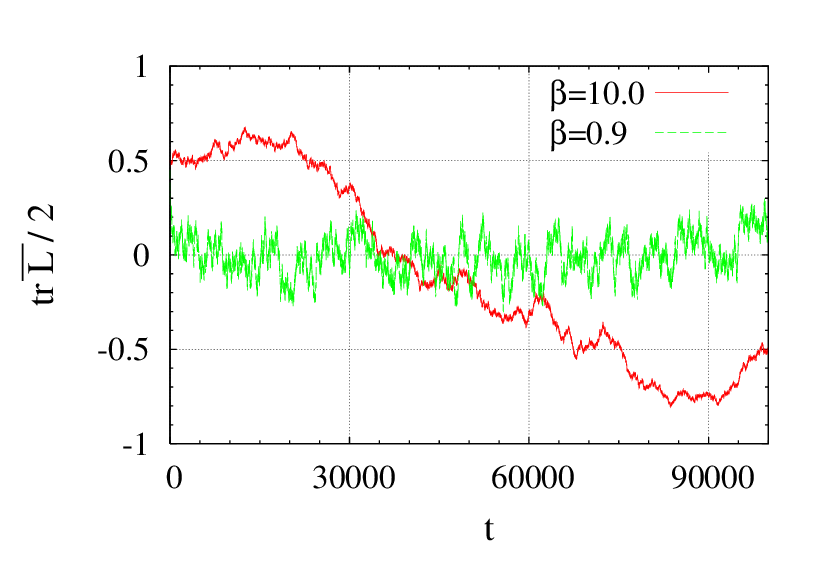}
\caption{Time evolution of the volume-averaged Polyakov loop below and
  far above $\beta_c$; $N_s=12$ lattice.
\label{fig:KSS_trL_t}}
\end{figure}
In Fig.~\ref{fig:KSS_trL_t} we show two time sequences for the
Polyakov loop just below and far above $\beta_c$. It is clear that
below the phase transition there are only small fluctuations about 0,
which decrease on larger lattices. On the other hand, at large
$\beta$, the Wilson lines partly align and $|\overline\ell|$ is far
from 0 for long time intervals. However, the above-mentioned slow rotation
of $\overline\boldl$ in group space (in the absence of a background
field) is clearly visible. We emphasize that Fig.~\ref{fig:KSS_trL_t}
depicts two particular runs which were much shorter than those used
for measurement.

Next, we consider two-point matrix-matrix correlation functions of the
form
\be \label{eq:CorFunc}
{\cal C}_\boldl(r) = \frac{1}{3} \frac{1}{N_s^3}
     \sum\limits_{\hat{\bm r}, \bm{r}_0} \frac{1}{2} \left< 
       \tr\boldl^\dagger(\bm{r}_0) \boldl(\bm{r}_0+\bm{r})\right>~.
\ee
The vector $\bm{r}$ is allowed to point in any of the three principal
directions of the lattice (in the positive direction only), over which
we average. Also, its length is restricted to $< N_s/2$ due to the
periodic boundary conditions.

Having determined the two-point function ${\cal C}(r)$ and its
statistical error, we perform a $\chi^2$ fit to the functional form
\be \label{eq:FitCorLength}
{\cal C}_\boldl(r) \sim \frac{1}{r m_\xi} \; e^{-r m_\xi} ~+ {\rm const.}
\ee
to extract the inverse spatial correlation length $m_\xi$. The
fits were restricted to $r\ge4$ (in lattice units) such that
$\chi^2/{\rm dof}\simeq1$.

\begin{figure}
\includegraphics*[width=\linewidth]{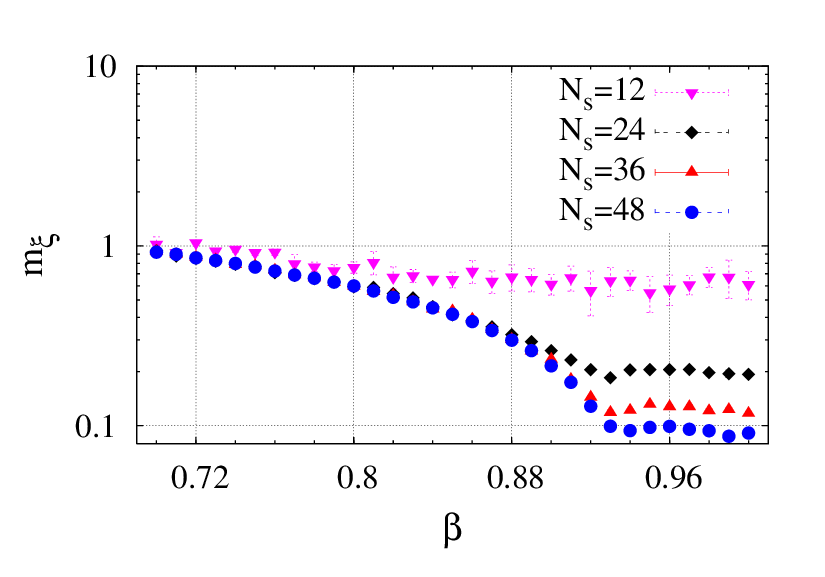}
\caption{The inverse spatial correlation length as a function of the coupling
  $\beta$ for various lattices.
\label{fig:KSS_xi}}
\end{figure}
Fig.~\ref{fig:KSS_xi} displays $m_\xi(\beta)$ for lattices of various
sizes. Deep in the disordered phase correlations extend only over a
few lattice sites and $m_\xi$ is therefore independent of the
volume. This confirms that both the $\tilde\ell^a$ and the Polyakov
loop $\ell$ are massive.  Long-range correlations do develop near
$\beta_c$ and $m_\xi$ drops to nearly zero, up to finite-size
effects. A fit of the form $m_\xi(\beta_c)\sim N_s^{-1/\nu_\xi}$
($N_s=24$, 36, 48 lattices only) gives the scaling exponent
\be
\nu_\xi = 0.938(5)~.
\ee
Quite clearly, there are massless modes (again, up to finite-size
effects) even above $\beta_c$ and hence $m_\xi$ remains small. These
observations are in line with the behavior of the integrated
autocorrelation time $\tau_{\rm int}$ for $\beta>\beta_c$ mentioned
above. We have also measured the correlation lengths for Polyakov
loops and for the adjoint $\tilde\ell^a$ fields via fits of the
form~(\ref{eq:FitCorLength}) to the two-point functions
\bea
{\cal C}_\ell(r) &\sim&
    \sum\limits_{\hat{\bm r}, \bm{r}_0} \left< 
       \ell(\bm{r}_0)~ \ell(\bm{r}_0+\bm{r})\right>~, \\
{\cal C}_{\tilde\ell}(r) &\sim&
    \sum\limits_{\hat{\bm r}, \bm{r}_0} \left< 
       \tilde\ell(\bm{r}_0)\cdot \tilde\ell(\bm{r}_0+\bm{r})\right>~.
\eea
We refrain from showing the results here since they closely resemble
$m_\xi(\beta)$ from Fig.~\ref{fig:KSS_xi}. The fact that the
correlation length for $\ell$ appears to diverge even above $\beta_c$
is probably due to mixing with the Goldstone modes.

Finally, we determine the distribution of eigenvalues of the Wilson
lines. For any given configuration (i.e.\ Metropolis time $t$), we
compute the eigenvalues $\lambda_1$ and $\lambda_2$ of the Wilson
lines $\boldl$ at each lattice site. We introduce their difference and
average,
\bea
\rho_1(t,\bm{x}) &=& \frac{1}{2}\left|\lambda_1(t,\bm{x}) -
                             \lambda_2(t,\bm{x})\right|~~,~~ \nonumber\\ 
\rho_2(t,\bm{x}) &=& \frac{1}{2}\left|\lambda_1(t,\bm{x}) + 
                             \lambda_2(t,\bm{x})\right|~.
\eea
The ensemble of $\rho_1(t,\bm{x})$ defines its probability
distribution $P_1(\rho_1)$, and similarly for $P_2(\rho_2)$. These
can be turned into effective potentials for the sum and difference of
eigenvalues, respectively, via
\be \label{eq:Veff}
V_{\rm eff}(\rho_1) = -\log P_1(\rho_1)~~,~~
V_{\rm eff}(\rho_2) = -\log P_2(\rho_2)~.
\ee

\begin{figure}
\includegraphics*[width=\linewidth]{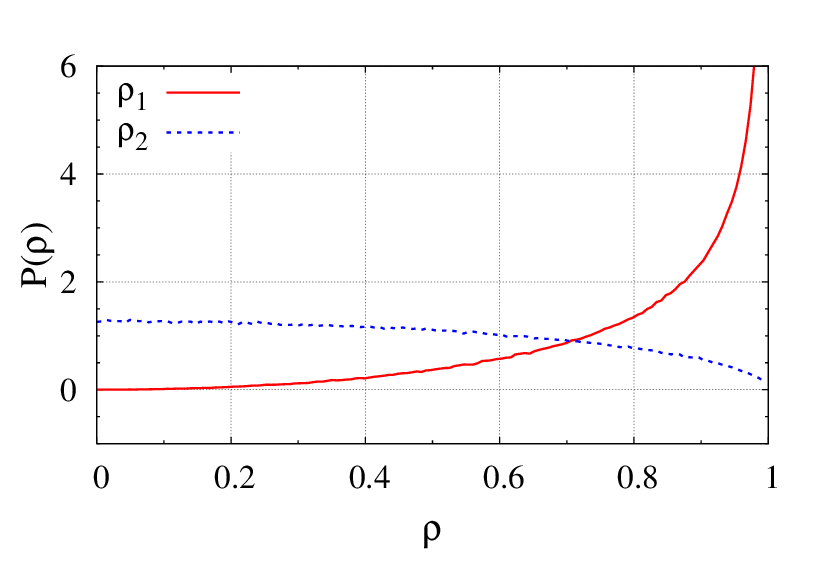}
\caption{The probability distributions of the difference ($\rho_1$)
  and sum ($\rho_2$) of eigenvalues of the Wilson line for $\beta=1$
  obtained on a $N_s=48$ lattice.
\label{fig:Prho_beta}}
\end{figure}
\begin{figure}
\includegraphics*[width=\linewidth]{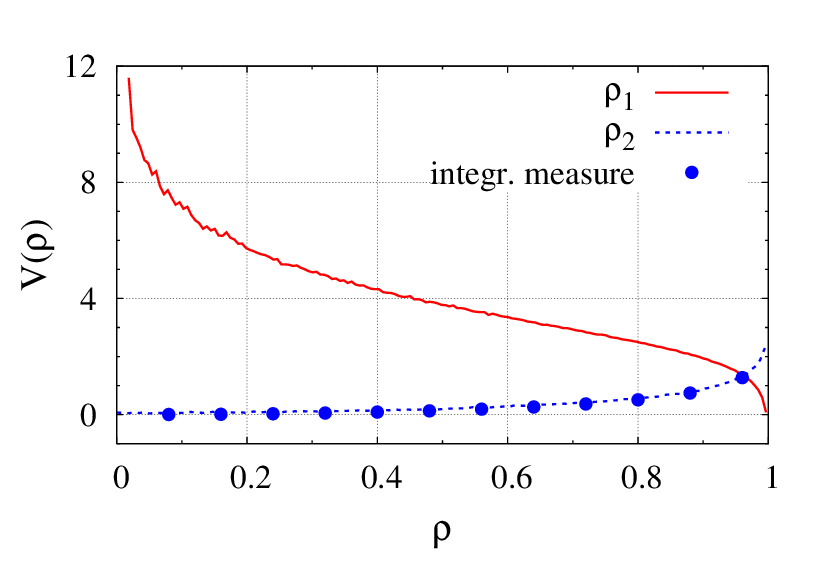}
\caption{The effective potentials for the difference ($\rho_1$) and
  sum ($\rho_2$) of eigenvalues of the Wilson line for $\beta=1$
  obtained on a $N_s=48$ lattice. The pure integration measure in
  terms of $\rho_2$, which is given by $\log (1-\rho_2^2)^{-1/2}$, is
  shown by the points.
\label{fig:Vrho}}
\end{figure}
Figures~\ref{fig:Prho_beta} and~\ref{fig:Vrho} depict the probability
distributions $P(\rho_1)$, $P(\rho_2)$ and the corresponding effective
potentials for $\beta=1$.  We have also determined these quantities
below the transition ($\beta=0.5$) but obtained very similar
curves. The potential shows evidence for a logarithmic divergence at
$\rho_1\to 0$ (or $\rho_2\to 1$); this is expected as the group
integration measure leads to logarithmic repulsion of the eigenvalues,
compare to eqs.~(\ref{eq:Weyl},\ref{eq:Fss_MF}). Aside from the
effects of the Vandermonde determinant, however, the eigenvalue
distribution (or the potential) for $\rho_2\equiv (1/2) |\tr \boldl|$
is entirely flat. This is illustrated in Fig.~\ref{fig:Vrho} which
compares the pure Vandermonde potential $\log (1-\rho_2^2)^{-1/2}$ to
the actually measured $V(\rho_2)$. The flat eigenvalue distribution is
consistent with the free global rotations of $\overline\boldl$
observed above.

\subsection{Action with ${\rm SU}(2)$ symmetry}
In this section, we add a mass term for the Polyakov loop $\ell= \tr
\boldl /2$,
\be  \label{eq:action_m2}
\cals = - \frac{1}{2}\beta\sum\limits_{\langle ij\rangle}
              \tr\left( \boldl_i\boldl_j^\dagger + {\rm h.c.}\right)
        - m^2 \sum\limits_i |\tr\boldl_i|^2~,
\ee
which explicitly breaks ${\rm SU}_L(2)\times {\rm SU}_R(2)$ to ${\rm
  SU}(2)$, $\boldl\to {\bf \Omega}^\dagger \boldl {\bf \Omega}$, and
also respects the Z(2) symmetry for the Polyakov loop, $\ell
\to-\ell$.  We study the phase structure as a function of $m^2$ at
fixed $\beta$. The order parameter for the deconfining phase
transition is given by the ensemble and volume averaged Polyakov loop
$\langle |\overline \ell|\rangle = \langle|\tr \overline
\boldl/2|\rangle$, where $\overline\boldl$ is defined in
eq.~(\ref{eq:KSSorder})\footnote{Taking the absolute value of
  $\overline \ell$ before performing the ensemble average is required
  due to the Z(2) symmetry.}. In this section, $\langle|\overline
\ell|\rangle$ will also be denoted as $\ell_0$.

\begin{figure}
\includegraphics*[width=\linewidth]{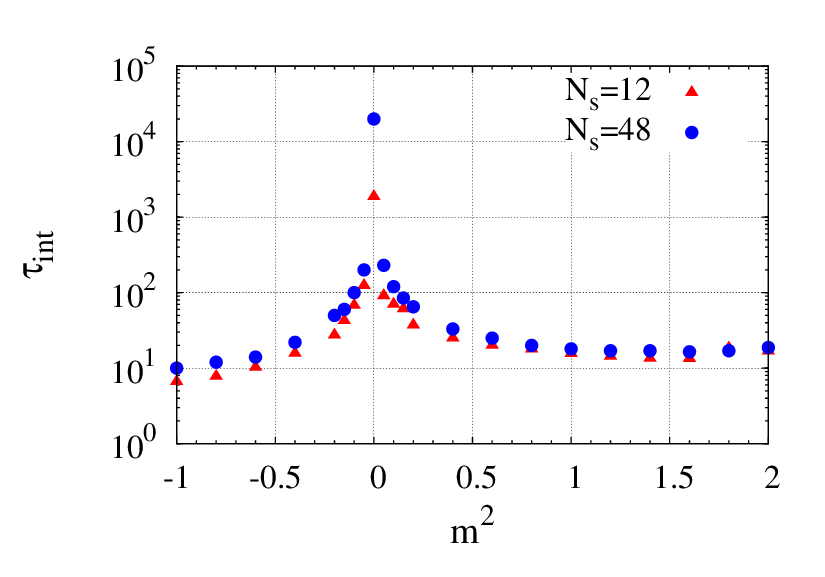}
\caption{The integrated autocorrelation time at $\beta=1$
as a function of the coupling $m^2$ for various lattices.
  \label{fig:m2_tauint}}
\end{figure}
In Fig.~\ref{fig:m2_tauint} we show the integrated autocorrelation
time for the Polyakov loop in the model~(\ref{eq:action_m2}) at
$\beta=1$, as a function of $m^2$. It indicates that the transition
occurs in the vicinity of $m^2\simeq0$, where $\tau_{\rm int}$ grows
with the lattice volume (critical slowing down). However, contrary to
Fig.~\ref{fig:KSS_tauint_beta}, above the transition point $\tau_{\rm
int}$ is independent of the volume. This confirms to our expectation
that long-range correlations in Metropolis-time should not appear for
$m^2\neq0$. A fit of the form $\tau_{\rm int}\sim N_s^{1/\nu_\tau}$ to
the $N_s=24$, 36, 48 data gives
\be
\nu_\tau = 1.3(4)~,
\ee
at $m^2=0$.  As before, all subsequent measurements were performed
with configurations that were separated by a time interval of
$\tau_{\rm int}$ (at least).

\begin{figure}
\includegraphics*[width=\linewidth]{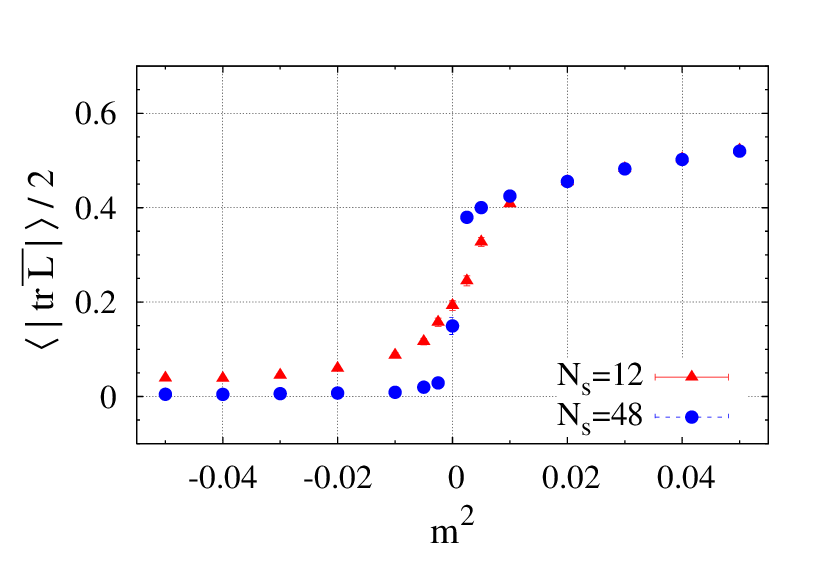}
\caption{The expectation value of the Polyakov loop as a function
  of the coupling $m^2$ (at $\beta=1$) for various lattices.
\label{fig:ell0_m2}}
\end{figure}
\begin{figure}
\includegraphics*[width=\linewidth]{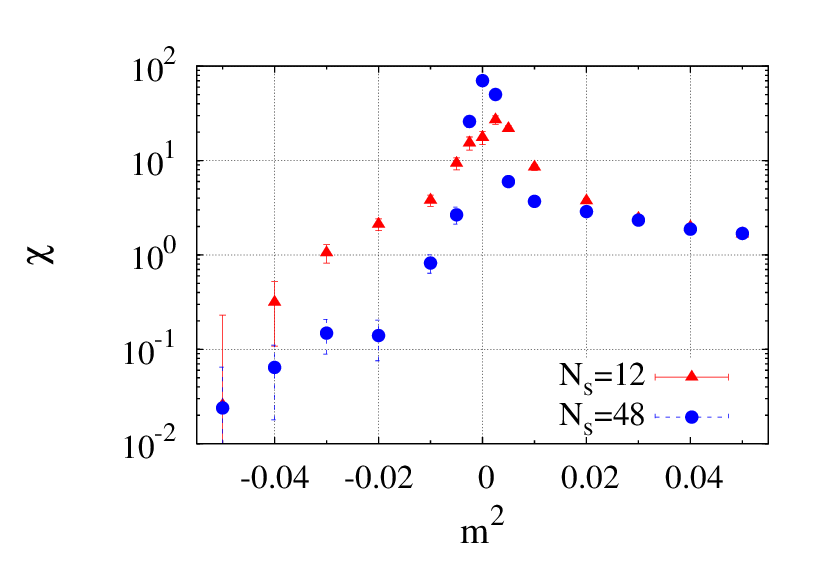}
\caption{The derivative of the Polyakov loop
  with respect to $m^2$ as a function of the
  coupling $m^2$ (at $\beta=1$) on various lattices.
\label{fig:chi_m2}}
\end{figure}
\begin{figure}
\includegraphics*[width=\linewidth]{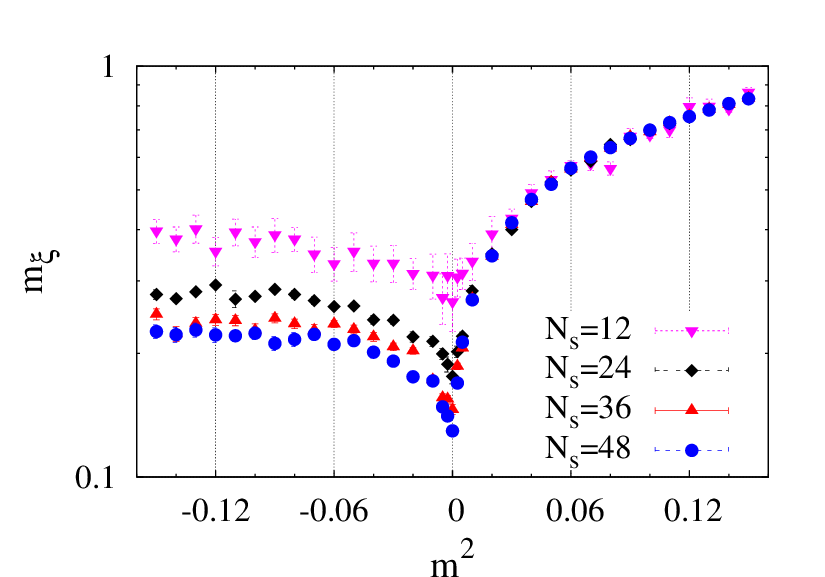}
\caption{The inverse spatial correlation length as a function of the coupling
  $m^2$ (at $\beta=1$) for various lattices.
\label{fig:m2_xi}}
\end{figure}

Figs.~\ref{fig:ell0_m2} and~\ref{fig:chi_m2} show the expectation
value of the Polyakov loop, and its derivative with respect to the
coupling, in a narrow window about the deconfining phase
transition. Within errors, we find that the critical coupling is
\be
m^2_c = 0.000(2)~.
\ee
The transition in terms of $m^2$ is evidently rather
sharp. Nevertheless, the scaling of $\tau_{\rm int}$ with the lattice
size mentioned above suggests a second-order phase transition in
infinite volume. This is confirmed also by the behavior of the
inverse correlation length $m_\xi(m^2)$ shown in Fig.~\ref{fig:m2_xi}.
$m_\xi$ has been determined by the same procedure outlined in
eqs.~(\ref{eq:CorFunc},\ref{eq:FitCorLength}) from the previous
section, and appears to vanish at $m^2=0$, $N_s\to\infty$; fitting
$m_\xi\sim N_s^{-1/\nu_\xi}$ (to the $N_s=24$, 36, 48 data) gives the
scaling exponent
\be
\nu_\xi = 2.28(8)~.
\ee
In the deconfined phase at $m^2>0$, the correlation length decreases
rapidly to about one (in lattice units). It decreases also as one
goes to negative values of $m^2$, into the confined phase, but
less rapidly. There, a weak volume dependence remains even from
$N_s=36$ to $N_s=48$.

\begin{figure}
\includegraphics*[width=\linewidth]{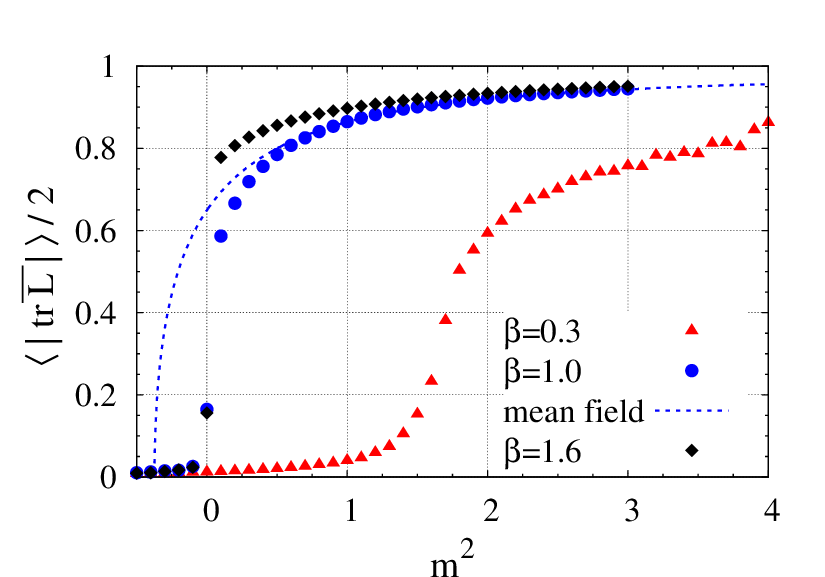}
\caption{The expectation value of the Polyakov loop as a function of
  the coupling $m^2$ at various $\beta$. Monte-Carlo results obtained
  on a $N_s=12$ lattice are indicated by the symbols. The line
  indicates the mean-field prediction for $\beta=1$, shifted
  horizontally by $m^2_{\rm mf} = m^2 - 0.94$. \label{fig:ell0_mf_m2}}
\end{figure}
The expectation value of the Polyakov loop is shown again in
Fig.~\ref{fig:ell0_mf_m2} over a broader range of $m^2$.  We also
compare to the mean-field prediction (only for $\beta=1$) discussed in
section~\ref{sec:MFA}, which has been shifted to the right by $\Delta
m^2=0.94$ to match the data far above the transition. Such a shift is
expected by analogy to the tadpole contribution in a scalar theory,
for example.  Not surprisingly, mean-field theory works well for large
$|m^2|\simge 0.5$ (far from the transition, to both sides), when the
effective masses are large and fluctuations are suppressed. Close to
the phase transition, critical fluctuations invalidate the mean-field
approximation.

The transition becomes extremely sharp when $\beta$ is large,
switching almost instantly from the confined phase to a perturbative
deconfined phase with $\ell_0\simeq1$. This behavior is in line with
the discussion in sections~\ref{sec:Intro} and~\ref{sec:LattAct}:
positive $m^2$ and large $\beta$ corresponds to the weak-coupling
limit of the four-dimensional theory in the deconfined phase. It can
also be readily understood from the expression~(\ref{eq:action_m2})
for the action: at large $\beta$ the Wilson lines at neighboring sites
are forced to align such that $\tr \boldl_i \boldl_j^\dagger/2
\approx1$. The potential only determines the direction of alignment:
when $m^2>0$, the preferred direction is the unit matrix (eigenvalue
attraction); when $m^2<0$, the Wilson lines instead live in the
subspace spanned by the Pauli matrices (eigenvalue repulsion).

Fig.~\ref{fig:ell0_mf_m2} also shows that the deconfining phase
transition is shifted to $m^2>0$ when $\beta<\beta_c$. In this limit
the alignment of the Wilson lines is enforced by the upside-down
potential rather than the nearest-neighbor interaction.

\begin{figure}
\includegraphics*[width=\linewidth]{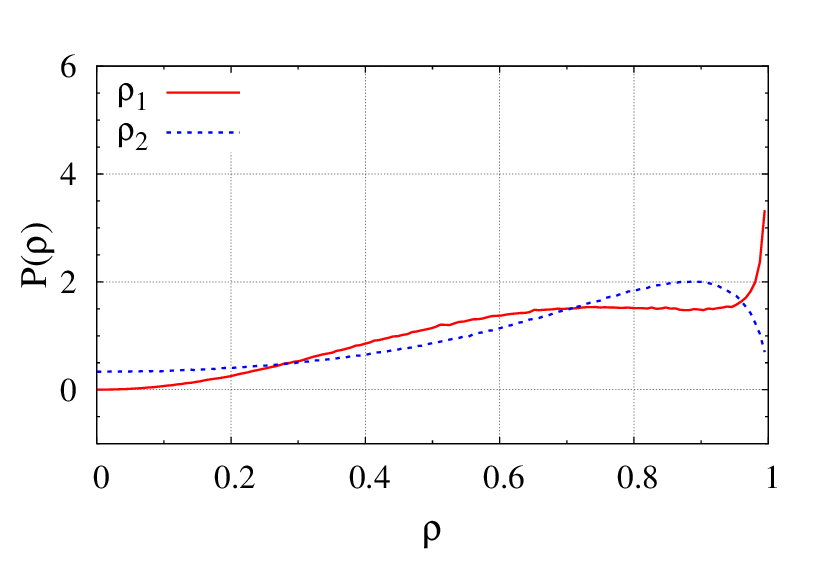}
\caption{Eigenvalue distributions for $m^2=0.15$ and $\beta=1$;
  $N_s=48$ lattice.
\label{fig:Prho_m2_015}}
\end{figure}
Fig.~\ref{fig:Prho_m2_015} depicts the eigenvalue distribution at
$\beta=1$ and $m^2=0.15$ which exceeds the critical $m_c^2$ for
deconfinement (since $\ell_0\simeq0.6$) but is still far from
asymptotic. Here, the perturbative potential~(\ref{Veff_1loop}) is
partly cancelled by the ``fuzzy bag'' term~(\ref{Veff_fuzzyB}) and the
eigenvalue distributions are rather broad. This result demonstrates
that the ``fuzzy bag'' term can generate eigenvalue repulsion in the
deconfined phase, in the regime $\beta\approx\beta_c$ corresponding to
moderately weak coupling in the underlying four-dimensional theory.

\begin{figure}
\includegraphics*[width=\linewidth]{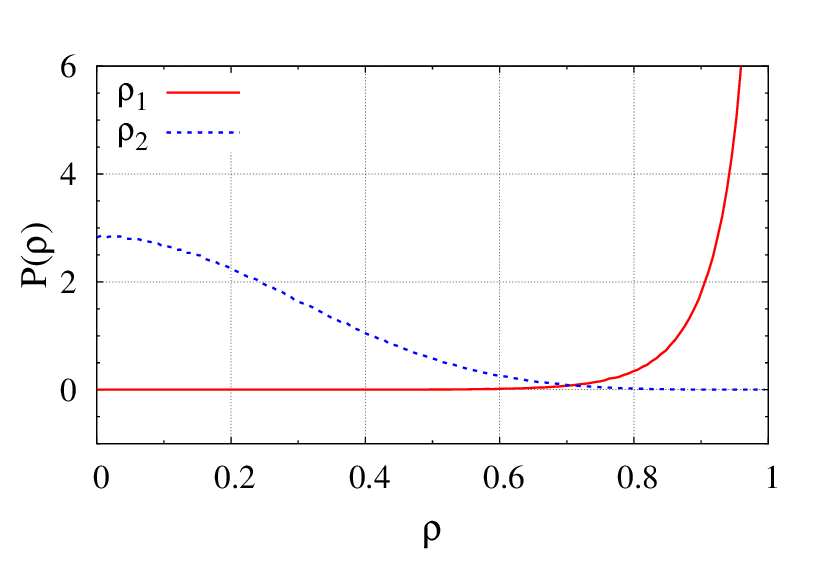}
\caption{Eigenvalue distributions for $m^2=-0.05$ and $\beta=3$;
  $N_s=48$ lattice.
\label{fig:Prho_beta3_m2_n005}}
\end{figure}

In the confined phase at $m^2<0$ and $\beta>\beta_c$ the Wilson
lines fluctuate about the non-trivial vacuum $\boldl_c=i\tau_3$, or
SU(2) rotations
thereof~\cite{Pisarski:2006hz,Meisinger,Schaden:2004ah}; this was
shown in Fig.~\ref{fig:Prho_beta} already. As expected, the
fluctuations diminish with increasing $\beta$, see
Fig.~\ref{fig:Prho_beta3_m2_n005}.
They are visible mostly in the distribution of the average eigenvalue
$\rho_2$ while $P(\rho_1)$ is rather sharp. This can be understood
easily by parameterizing the fluctuations about $i\tau_3$ as
$\boldl\sim i\,{\rm diag}\,(e^{i\phi},-e^{-i\phi})$, with
$\phi\approx0$. Then, \bea \rho_1 &=& |\cos \phi| \simeq
1-\frac{\phi^2}{2}~,\\ \rho_2 &=& |\sin \phi| \simeq |\phi|~.  \eea

\begin{figure}
\includegraphics*[width=\linewidth]{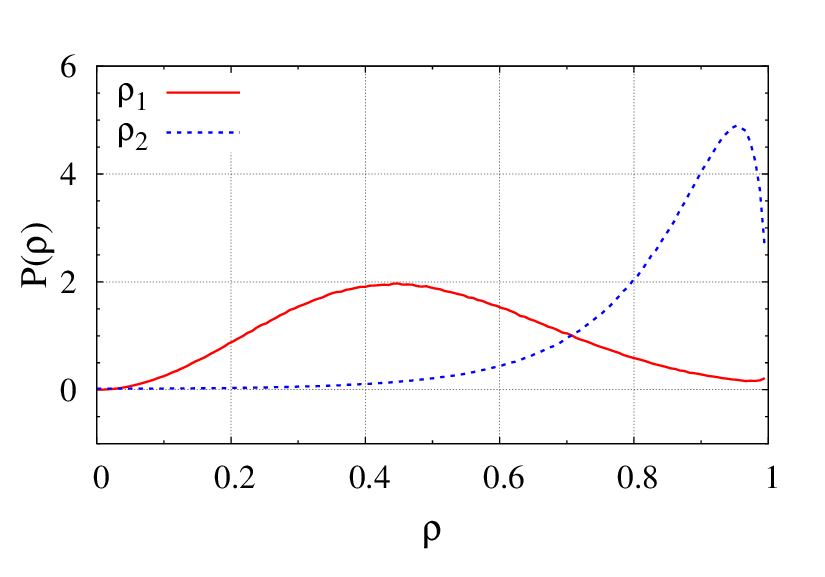}
\caption{Eigenvalue distributions for $m^2=0.8$ and $\beta=1$;
  $N_s=48$ lattice.
\label{fig:Prho_m2_08}}
\end{figure}
For large $m^2$ one of course approaches the perturbative vacuum, as
shown in Fig.~\ref{fig:Prho_m2_08}. The distribution for
$\rho_2=|\ell|$ peaks near 1 while that for the difference of
eigenvalues is broader. In the perturbative regime fluctuations can be
parameterized as $\boldl\sim {\rm diag}\,(\exp\,i\phi,\exp\,-i\phi)$,
with $\phi\approx0$. Hence, the fluctuations of $\rho_1 = |\sin\phi|$
are much bigger than those of $\rho_2 = |\cos\phi|\simeq 1-
\phi^2/2$.  The fact that $P(\rho_1)\to0$ as $\rho_1\to0$, and
$P(\rho_2)\to0$ as $\rho_2\to1$, is again due to the integration
measure, see eq.~(\ref{eq:Weyl}). For even larger $m^2$, both
distributions get sharper and their maxima move further towards
$\rho_1=0$ and $\rho_2=1$, respectively. In all, far above the
transition the eigenvalue distributions qualitatively exhibit the
behavior appropriate for the perturbative weak-field regime.

\section{Summary and Conclusions}

\begin{figure}
\includegraphics*[width=6cm,height=7cm]{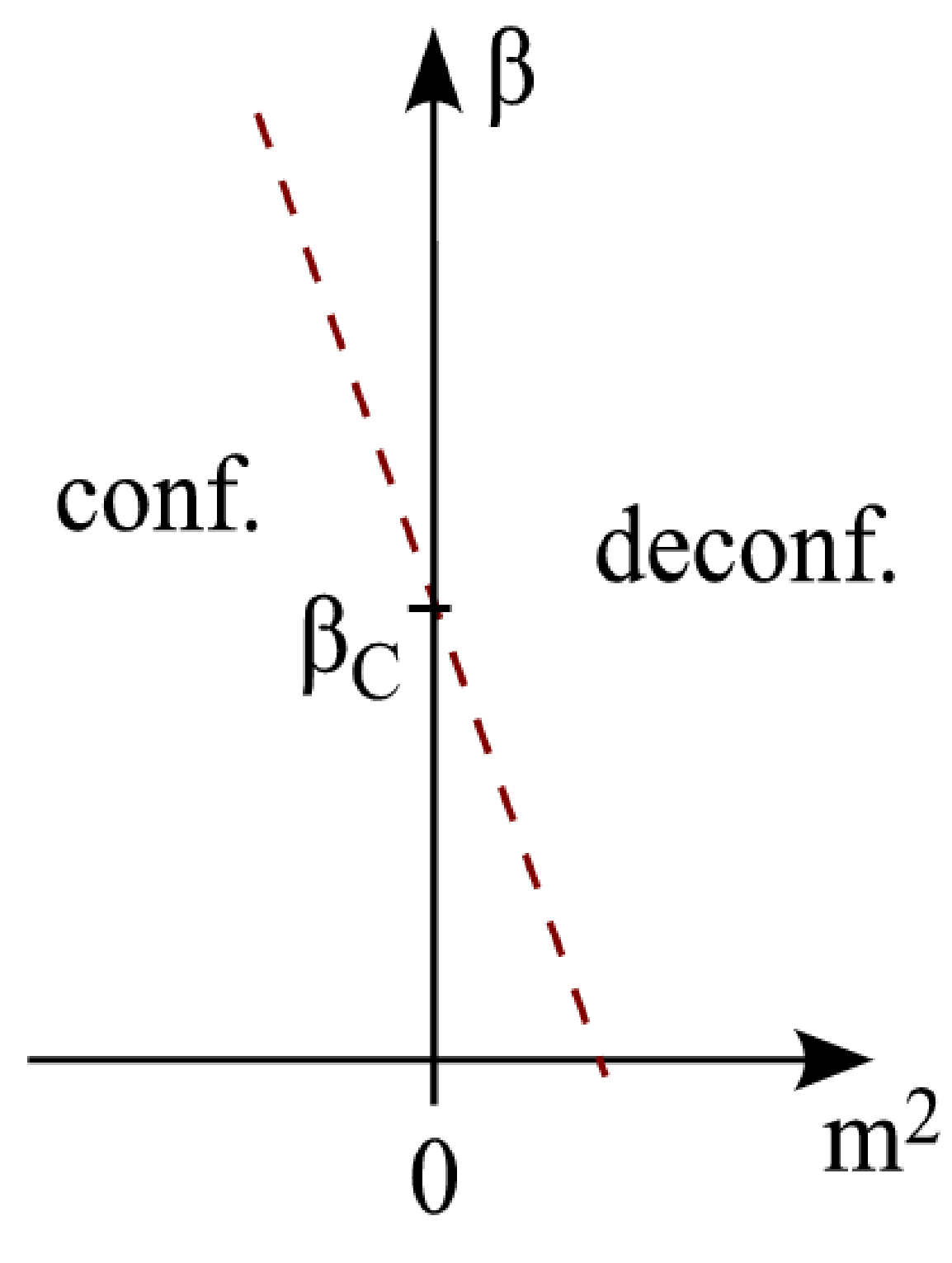}
\caption{Schematic sketch of the phase diagram in the $\beta-m^2$
  plane, for infinite volume. The presence of an infinitesimal
  background field $-h\,\tr\boldl$ is assumed.
\label{fig:PhaseDia}}
\end{figure}
We have performed Monte-Carlo simulations of an effective theory of
SU(2) Wilson lines in three dimensions. The main purpose of this work
was a study of eigenvalue repulsion in the deconfined phase of a SU(2)
matrix model. We considered the action
\be \label{Sum_Action}
\cals = - \frac{1}{2}\beta\sum\limits_{\langle ij\rangle}
              \tr\left( \boldl_i\boldl_j^\dagger + {\rm h.c.}\right)
        - m^2 \sum\limits_i |\tr\boldl_i|^2~,
\ee
without gauge fields, $A_i^a=0$. The kinetic term exhibits a global 
${\rm SU}_L(2)\times {\rm SU}_R(2)$ symmetry which is broken
explicitly to ${\rm SU}(2)$ by the loop potential. Note that a
$\boldl\to {\bf \Omega}_L \boldl {\bf \Omega}_R$
transformation changes the eigenvalues of $\boldl$, while $\boldl\to
{\bf \Omega}^\dagger \boldl {\bf \Omega}$ does not.

The phase diagram is sketched in Fig.~\ref{fig:PhaseDia}. In the
absence of a potential, at $m^2=0$, (\ref{Sum_Action}) is essentially a
standard spin-model. At small $\beta$ the effective mass of the Wilson
lines is large and they fluctuate independently from site to
site. Confinement is realized in a trivial way since $\overline\boldl
\to \bm{0}$ for each configuration, where $\overline\boldl$ denotes
the volume-averaged Wilson line. This remains true for small $|m^2|$.
To deconfine, a large upside-down potential ($m^2>0$) is required to
align the Wilson lines to the unit matrix. Hence, for small $\beta$
the phase transition arises due to the effective {\em loop potential},
in a regime where ${\rm SU}_L(2)\times {\rm SU}_R(2)$ is broken
strongly.

There is a second-order phase transition at $\beta_c\simeq0.942$ (and
$m^2=0$) where the masses (inverse correlation lengths) of the
Polyakov loop $\ell= \tr\boldl/2$ and of the adjoint fields $\tilde
\ell^a= -i \, \tr \boldl \bm\tau^a/2$ vanish. This is associated with
spontaneous breaking of ${\rm SU}_L(2)\times {\rm SU}_R(2)$ to
$SU(2)$, where three Goldstone modes appear. We have confirmed that
the ``length'' $u^2=\tr \overline\boldl^\dagger \overline\boldl/2$ of
$\overline\boldl$ aquires a non-zero expectation value for $\beta
>\beta_c$. Hence, we expect that a weak background field
$-h\,\tr\boldl$, $h\to0$, shifts the phase boundary to $m^2<0$ as
indicated in Fig.~\ref{fig:PhaseDia}.

Very large lattice coupling $\beta\gg1$ corresponds to the extreme
weak-coupling limit of the original four-dimensional theory; the
effective theory can nevertheless confine because it incorporates the
global Z(N) symmetry for the Polyakov loop. At large $\beta$
fluctuations are suppressed and the Wilson lines are again forced to
align, this time by the nearest-neighbor interaction (kinetic
term). The direction of alignment is determined by the loop
potential. A standard potential with positive curvature ($m^2<0$) is
minimized by Wilson lines with no singlet component, hence eigenvalues
repel and the theory confines\footnote{We expect that the phase
  boundary is shifted from $m^2=0$ to some smaller value if an
  infinitesimal background field is applied.}. On the other hand, an
upside-down potential ($m^2>0$) leads to $\boldl(\bm{x}) \sim \dblone$
and so to eigenvalue attraction and deconfinement. For $\beta\gg1$
even a weak potential suffices to trigger the locking into (or out of)
the center of the group. This leads to a sharp transition directly to
a perturbative deconfined phase without eigenvalue repulsion.

We have measured the distributions of the eigenvalues of the Wilson
line in the non-perturbative deconfined phase above, but close to,
$\beta_c$. They show clearly the emergence of eigenvalue repulsion
even for ``temperatures'' (i.e.\ $m^2$) not extremely close to the
phase boundary. It is only relatively deep in the deconfined phase
($m^2\simge1$) that the distribution of eigenvalues peaks near 1,
which corresponds to the perturbative vacuum. These results confirm
the suggestion of ref.~\cite{Pisarski:2006hz} that eigenvalue
repulsion in the deconfined phase does arise at intermediate values of
the nearest-neighbor coupling $\beta$, due to fluctuations of the
Wilson lines, provided that the non-perturbative ``fuzzy-bag'' term
approximately cancels the perturbative loop potential. Such a ``fuzzy
bag'' contribution in the effective theory makes it possible to reach
the region of small $m^2$ in the phase diagram.

In the confined phase at $\beta>\beta_c$ the volume-averaged Wilson
line $\overline\boldl$ approaches the center-symmetric
vacuum~\cite{Pisarski:2006hz,Meisinger,Schaden:2004ah}
\be \label{eq:Vac_Lc}
\boldl_c={\rm diag}\; (1,z,z^2,\cdots,z^{N-1})~~~,~~~(z\equiv e^{2\pi i/N})~,
\ee
which for two colors corresponds to $\boldl_c=i\,\tau_3$ (up to an
overall SU(2) rotation). This is due to the fact that the
Wilson lines align at large $\beta$, and $m^2<0$ favors a direction
orthogonal to unity. We repeat that this is not the case when $\beta$
is small, where instead $\overline\boldl\to0$ for $m^2\simeq0$.

\section*{Acknowledgments}
We are indebted to Rob Pisarski for many helpful
discussions. D.S.\ gratefully acknowledges a fellowship by the
Helmholtz foundation. The numerical simulations presented
here were performed at the Center for Scientific Computing (CSC) at
Frankfurt University. Our code is based in part on the MILC
collaboration's public lattice gauge theory code, see\\
http://physics.utah.edu/\~{}detar/milc.html.

\end{document}